\documentclass[prb,twocolumn]{revtex4}

\usepackage{graphicx}
\usepackage{amsmath}
\usepackage{amssymb}
\usepackage{latexsym}

\begin{document}

\title{Quantum antidot as a controllable spin injector and spin filter}
\author{I. V. Zozoulenko and M. Evaldsson}
\affiliation{Department of Science and Technology (ITN), Link\"oping University, 601 74
Norrk\"oping, Sweden}
\date{\today}

\begin{abstract}
We propose a device based on an antidot embedded in a narrow
quantum wire in the edge state regime, that can be used to inject
and/or to control spin polarized current. The operational
principle of the device is based on the effect of resonant
backscattering from one edge state into another through a
localized quasi-bound states, combined with the effect of Zeeman
splitting of the quasibound states in sufficiently high magnetic
field. We outline the device geometry, present detailed
quantum-mechanical transport calculation and suggest a possible
scheme to test the device performance and functionality.
\end{abstract}

\maketitle

The most ambitious and challenging long-term aim of
semiconductor-based spintronics is the practical implementation of
quantum information processing based on the spin properties of the
electron.  One of the most promising practical realization areas
of quantum logic are low dimensional semiconductor structures like
quantum dots and related lateral structures defined in a
two-dimensional electron gas. Two coupled dots, for example, can
serve as the simplest two-qbit quantum gate\cite{DiVincenzo}.

There has been several suggestions on how to implement the spin
control in quantum dots and related systems\cite{proposals}, and
some of the proposed schemes has already demonstrated their
potential for injection and detection of spin-polarized current
\cite{Marcus}. The spin-polarized injection and detection of the
electrons in a quantum dot has recently been demonstrated in the
edge state regime \cite{Andy_PRL2002}. However, in the above
experiment no control over the spin state was possible (it was
always spin-down spins that were injected into the dot).

In a present paper we propose a new method to control the spin
freedom in the edge state regime  by making use of an
\textit{antidot} embedded in a narrow quantum wire. Based on the
proposed device, a spin population  can be achieved in a
\textit{controlled} way, and the read-out spin information can be
converted into transport properties. In the present paper we
outline the device geometry, present detailed quantum-mechanical
transport calculation and suggest a possible scheme to test the
device performance and functionality.

Quantum antidot is a potential hill defined in 2DEG by means of
e.g. electrostatic split
gates\cite{Andy_exp,Andy_exp_theor,Goldman,Cambridge_1994,
Cambridge_1995,Cambridge_2000,Cambridge_2003}. In a perpendicular
magnetic field electrons are trapped around the antidot in a bound
state that is formed due to the magnetic confinement. In a
classical picture this corresponds to electron skipping orbits
around the antidot with a classical cyclotron radius $r_c$. The
operational principle of the proposed device is based on the
effect of resonant backscattering from one edge state into another
through localized quasi-bound states
\cite{Kivelson,Andy_exp_theor,theor}, combined with the effect of
Zeeman splitting of the quasibound states in sufficiently high
magnetic field.

\begin{figure}[!htp]
\includegraphics[scale=0.6]{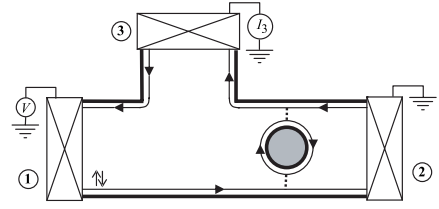}
\caption{\label{fig1} Schematic geometry of the device. An antidot
is defined in one arm of a quantum wire in a three terminal
geometry. The wire supports two spin-resolved channels (i.e. the
lowest edge state with spin-up and spin-down electrons).}
\end{figure}

The geometry of the device is presented in Fig. \ref{fig1}. An
antidot is defined in one arm of a quantum wire in a
three-terminal geometry. Leads 2 and 3 are kept at the same
voltage (e.g. grounded) such that there is no current flow between
them. We assume that magnetic field is sufficiently high allowing
only two spin-resolved channels.
 The system is described by
Hamiltonian  $H=\sum_\sigma H_\sigma$, that in the Landau gauge
$\mathbf{A}=(-By,0)$ reads
\begin{eqnarray}\label{Hamiltonian}
H_\sigma=\frac{\hbar^2}{2m^*} \left[
\left(\frac{\partial}{\partial x}
-\frac{ieBy}{\hbar}\right)^2+\frac{\partial^2}{\partial
y^2}\right]+V(x,y)+g\mu_B\sigma B
\end{eqnarray}
where $m^*=0.067m_e$ is the effective electron mass for GaAs,
$V(x,y)$ is the external confining potential. The last term in
Eq.~(\ref{Hamiltonian}) accounts for Zeeman energy where
$\mu_B=e\hbar/2m_e$ is the Bohr magnetron, $\sigma=\pm\frac{1}{2}$
describes spin-up and spin-down states, $\uparrow ,\downarrow$,
and the $g$ factor of GaAs is $g=-0.44$. For the case of
phase-coherent electron dynamics, the transport through device is
described within the Landauer-B\"uttiker formalism that relates
the conductance of the multi-terminal device to its scattering
properties \cite{Buttiker}. In a sufficiently high magnetic field
there is no overlap between the edge states running along opposite
sides of the channel, such that $T_{23}=T_{12}=T_{11}=0,
T_{13}=1$, where $T_{ji}$ defines the transmission probability
from lead $i$ to lead $j$. Define the transmission coefficients of
the system as a fraction of electrons that are injected from lead
1 and after being backscattered by the antidot are transmitted
into lead 3,
\begin{eqnarray}\label{TM}
T\equiv T_{31}
, \end{eqnarray}
 (note $T_{31}+T_{21}=N$, where the number of spin-resolved channels $N=2$).
 Using the Landauer-B\"uttiker formula we obtain that the current
 flowing out of
the lead 3 is simply determined by the transmission coefficient
$T$
\begin{eqnarray}\label{G}
I_3=GV,
\ G=\left(e^2/h\right)T. \end{eqnarray}
In the absence of the Zeeman term the conductance $G=I_3/V$ would
exhibit a series of pronounced peaks whenever the Fermi energy
$E_F$ matches the antidot resonant state energies\cite{Kivelson}.
In the presence of the Zeeman splitting, the energy of each
quasibound state is different for electrons of the opposite spins.
This leads to the splitting of the corresponding conductance
peaks. We thus expect that by varying parameters of the system
(e.g. magnetic field, the Fermi energy, the antidot size) the
conductance of the device can be tuned within each of the split
peaks such that the current in lead 3 will be due exclusively to
spin-up or spin-down electrons. Thus, the proposed device can
operate as a controllable spin injector. If the incoming state in
the lead 1 is already spin-polarized (say, only spin-ups are
injected into the lead 1), tuning the energy of the quasibound
state to the one corresponding to spin-down electrons would
suppress the current completely, i.e. the device would operate as
a spin filter or switcher.

In order to test feasibility of the proposed device we perform
full quantum mechanical transport calculations for the case of an
antidot system defined in GaAs heterostructure. In order to
calculate the transmission coefficient we use the ``hybrid"
recursive Green function technique (the details of the method can
be found in \cite{Z}).  We use the model of a hard wall
confinement to approximate the potential profile of the antidot
and the wire. This is certainly a simplification in comparison to
the actual smooth potential profile in the device. This model is
however fully sufficient for the present purposes
because in the one-electron picture the use of the smooth instead
of the hard wall confinement would only change a position and a
width of the conductance peaks, but would not affect the main
features of the conductance.

\begin{figure}[!htp]
\includegraphics[scale=1]{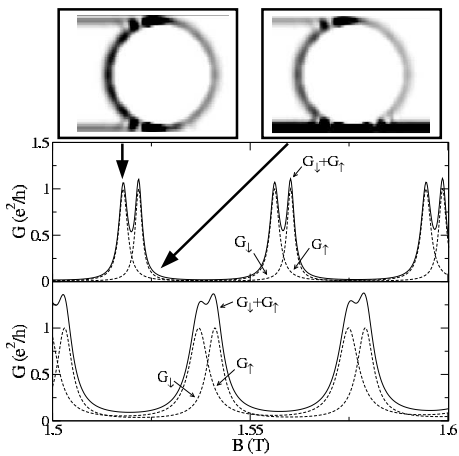}
\caption{\label{fig2} The conductance of the device $G=I_3/V$ as a
function of perpendicular magnetic field $B$. The middle and lower
panels
correspond to the cases of weak and strong coupling where the
antidot diameter is $d=318$ nm and $d=320$ nm respectively. The
width of the wire $w=400nm$, and the sheet electron density
$n_s=10^{15}m^{-2}$. The upper panel shows the current density for
spin-up electrons for the resonant and off-resonant transmission.}
\end{figure}

Figure \ref{fig2} shows the three-terminal conductance of the
device $G=G^\uparrow +G^\downarrow$ as a function of magnetic
field, where $G^\uparrow$ and $G^\downarrow$ represent the
conductance of the individual spins channels given by Eq.
(\ref{G}). The conductance $G=G(B)$ exhibits periodic
oscillations, where the periodicity (i.e. the distance between two
pairs of the split peaks) is related to the addition of one flux
quantum $\phi_0=h/e$ to the total flux $\Phi=BS$ ($S$ being the
area of the antidot). This gives $\Delta B=h/eS=0.04$T which is
consistent with the calculated periodicity $\Delta B=0.05$T (note
that the area enclosed by the bound state is larger than $S$
because of the finite spatial extend on the wave function). In the
upper panel of Fig. \ref{fig2} we illustrate the effect of
resonant backscattering via the quasibound antidot state by
plotting the current density plots for the cases of on- and
off-resonant transmission. It is worth mentioning that the
operation of the device is conceptually similar to add-drop
filters and mode couplers in quantum optics and optical
communications\cite{add-drop}.

The broadening of the peaks in each pair is determined by the
coupling strength between the bound state and the edge states.
This is illustrated in Fig. \ref{fig2} for the cases of weak and
strong coupling where two peaks in each pair are respectively
separated or practically merged. By decreasing the coupling
strength the peaks can be made sufficiently narrow such that the
efficiency of the spin injection (that we define as $\xi
=2|G_\uparrow -G_\downarrow|/(G_\uparrow +G_\downarrow)$) can
theoretically reach 100\%. In practice this value may be reduced
by e.g. temperature broadening, asymmetry in coupling to upper and
lower edge states etc. For a given magnetic field the spitting of
the peaks is determined by the $g$-factor of the material and thus
can not be changed. We note with this respect that utilization of
the material with higher $g$ factor, e.g. InAs where $|g|=15$ will
apparently lead to much more pronounced peak splitting and thus
higher efficiency $\xi$.

\begin{figure}[!htp]
\includegraphics[scale=0.5]{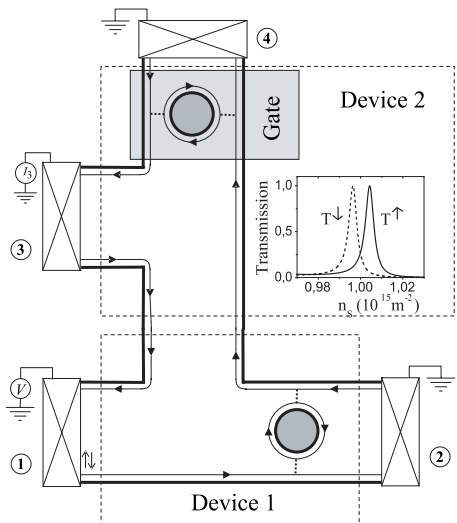}
\caption{Two antidot devices connected in series. A back gate at
device 2 varies locally the electron density $n_s$. The inset
shows the transmission coefficient $T=T^\uparrow + T^\downarrow$
of a single device as a function of the sheet electron density
$n_s=k_F^2/2\pi$ for the wire of the width $w=400$ nm and the
antidot diameter $d=318$ nm, $B=1.52$ T.\label{fig3}}
\end{figure}

Consider now two three-terminal devices connected in series as
shown in Fig. \ref{fig3}.  The edge state leaving the device 1 and
entering the device 2 is not reflected back to the device 1 (it
can leave through leads 3 or 4 only). Since multiple reflections
between two devices are absent, the transmission coefficient
$T_{31}^{4t}$ in a four-terminal geometry is simply given by the
sum $T_{31}^{4t}=T_1^\uparrow T_2^\uparrow + T_1^\downarrow
T_2^\downarrow$, where $T_1^{\uparrow(\downarrow)}$ and
$T_2^{\uparrow(\downarrow)}$ are the transmission coefficients of
individual devices 1 and 2 for spin-up (spin-down) states defined
by Eq. (\ref{TM}) (note that we neglect spin-flip processes such
that two spin channels are totally independent).

Let us now illustrate the operation of the device as a spin filter
or spin switcher. Consider the geometry of Fig. \ref{fig3} (two
devices connected in series).  Assume that parameters of the
device 2 can be independently changed, for example by a back gate
that varies locally the electron density $n_s$. The transmission
coefficient $T_2$ exhibits a series of split peaks as a function
of $n_s$, see inset to Fig. \ref{fig3}.
Assuming that leads 2 - 4 are grounded and applying
Landauer-B\"uttiker formalism we obtain for the four-terminal
conductance $G^{4t}=I_3/V=T_{31}^{4t}V$, where $I_3$ is the
current out of lead 3 and $V$ is the voltage at lead 1. Suppose
that magnetic field is tuned such that only one spin polarized
state (say spin-up) passes the device 1 and enters the device 2,
i.e. $T_1^\downarrow=0, T_1^\uparrow=1$. In this case
$T_{31}^{4t}=T_1^\uparrow T_2^\uparrow + T_1^\downarrow
T_2^\downarrow=T_2^\uparrow$. Thus, by varying the back gate
voltage, one can adjust $T_2^\uparrow$ such that the spin-up state
entering the device 2 is either totally transmitted or blocked.
Therefore, the proposed device operates as a controllable
filter/switch for incoming spin-polarized electrons.

We would like to stress that in our calculations we used a
one-electron Hamiltonian where electron-electron interaction is
effectively included in form of an external self-consistent
potential. This one-electron description was successfully used to
explain various features observed in a number of experiments on
antidot structures\cite{Andy_exp_theor,theor,Cambridge_1995}. It
has been argued however that charging effects may be important for
understanding of such features in the conductance as
double-frequency ($h/2e$) AB
oscillations\cite{Cambridge_1994,Cambridge_2000,Cambridge_2003}.
As magnetic field increases, the conductance peaks/dips (of the
periodicity $\Delta BS=h/e$ discussed above) split into pairs,
which was usually taken as the signature of Zeeman induced spin
splitting. As $B$ increases further, the splitting saturates such
that peaks/dips become equidistantly spaced (i.e. the oscillations
become $h/2e$-periodic). This effect was explained in
\cite{Cambridge_2000} in terms of charging of compressible regions
(CR) forming around antidots. Within this model the conductance
peaks/dips are due to electrons of the same spin tunnelling
through the outer spin states. It was however
noted\cite{Cambridge_2000} that it is unclear whether CR really
exist at low $B$ and it was suggested that a gradual transition
from conventional Zeeman splitting to the interacting picture may
take place as $B$ increases. Thus the detailed nature of $h/2e$
oscillations still remains an open question and its full
understanding requires accounting for spin and charging effect in
a self-consisted way. We currently undertake full
quantum-mechanical transport calculation combined with the
local-spin-density approximation in an attempt to answer this
question. Note that two antidot devices in series (functioning as
a spin filter/switch as described above) can be used to test
whether the $h/2e$ oscillations are due to electrons of the same
or different spins.

To conclude, we suggest an antidot-based device that can inject
and control spin-polarized current.
We hope that the present paper will stimulate further interest to
quantum antidot systems which provide a powerful tool to study,
control and manipulate the spin degrees of freedom of electrons.

We thank Andy Sachrajda for stimulating discussions and critical
reading of the manuscript. Financial support from Swedish Research
Council is greatly acknowledged. M.E. acknowledges a support from
National Graduate School in Scientific Computing.



\begin{thebibliography}{99}


\bibitem{DiVincenzo} D. Loss and D. P. DiVincenzo,
 Phys. Rev. A \textbf{57}, 120 (1998).

\bibitem{proposals}P. Recher, E. V. Sukhorukov, and D. Loss, Phys.
Rev. Lett. \textbf{85}, 1962 (2000); D. Frustaglia, M. Hentschel,
and K. Richter, Phys. Rev. Lett. \textbf{87}, 256602 (2001); M.
Val\'{i}n-Rodr\'{i}gues \textit{et al.}, Phys. Rev. B \textbf{66},
165302 (2002).

\bibitem{Marcus} R. M. Potok \textit{et al.}, Phys. Rev. Lett.
\textbf{89}, 266602 (2002).

\bibitem{Andy_PRL2002} P. Hawrylak \textit{et al.},
 Phys. Rev. B \textbf{59}, 2801 (1999);
M. Ciorga \textit{et al.}, Phys. Rev. Lett. \textbf{88}, 256804
(2002).

\bibitem{Andy_exp} Y. Feng  \textit{et al.}, Appl. Phys. Lett. \textbf{63}, 1666
(1993); C. Gould  \textit{et al.}, Phys. Rev. Lett. \textbf{26},
5272 (1996)

\bibitem{Andy_exp_theor}G. Kirczenow  \textit{et al.},
Phys. Rev. Lett. \textbf{72}, 2069 (1994); G. Kirczenow \textit{et
al.}, Phys. Rev. B \textbf{56}, 7503 (1997).

\bibitem{Goldman}I. J. Maasilta and V. J. Goldman, Phys. Rev. B
\textbf{57} R4273 (1998).

\bibitem{Cambridge_1994} C. J. B. Ford  \textit{et al.}, Phys. Rev. B \textbf{49},
17456 (1994)


\bibitem{Cambridge_1995} D. R. Mace  \textit{et al.}, Phys. Rev. B
\textbf{52}, R8672 (1995).

\bibitem{Cambridge_2000} M. Kataoka  \textit{et al.}, Phys. Rev. B
\textbf{62}, R4817 (2000).

\bibitem{Cambridge_2003}  M. Kataoka  \textit{et al.},
Phys. Rev. Lett \textbf{83}, 160 (1999) (2000); M. Kataoka,
\textit{et al.}, Phys. Rev. B \textbf{68} 153305 (2003);


\bibitem{Kivelson}J. K. Jain and S. A. Kivelson, Phys. Rev. Lett.
\textbf{60}, 1542 (1988).

\bibitem{theor} Y. Takagaki and D. K. Ferry, Phys. Rev. B
\textbf{48}, 8152 (1993); G. Kirczenow, Phys. Rev. B \textbf{50},
1649 (1994); Y. Takagaki,  Phys. Rev. B \textbf{55}, R16021
(1997); C. C. Wan, T. De Jesus, and Hong Guo, Phys. Rev. B
\textbf{57}, 11907 (1998).




\bibitem{Buttiker}M. B\"uttiker, Phys. Rev. Lett. \textbf{57}, 1761
(1986).


\bibitem{Z}I. V. Zozoulenko, F. A. Maa\o, and E. H. Hauge, Phys. Rev. B
\textbf{53}, 7975 (1996); I. V. Zozoulenko, F. A. Maa\o, and E. H.
Hauge, Phys. Rev. B \textbf{53}, 7985 (1996); I. V. Zozoulenko, F.
A. Maa\o, and E. H. Hauge, Phys. Rev. B \textbf{56}, 4710 (1997).


\bibitem{add-drop} S. Fan  \textit{et al.}, Phys. Rev. Lett.
\textbf{80}, 960 (1998); M. Cai,
O. Painter, and K. Vahala, Phys. Rev. Lett. \textbf{85}, 74
(2000).



\end{thebibliography}
\end{document}